\documentclass[prl,aps,twocolumn,amsmath,amsmath,amssymb,superscriptaddress]{revtex4-1}

\usepackage{graphicx}
\usepackage{dcolumn}
\usepackage{bm}
\usepackage[colorlinks,linkcolor=blue,citecolor=blue,urlcolor=blue]{hyperref}
\usepackage{float}
\usepackage[dvipsname]{xcolor}
\usepackage[utf8]{inputenc} 
\usepackage[T1]{fontenc}
\usepackage[normalem]{ulem}

\begin{document}

\title{\hbox{Ultrafast Hidden Spin Polarization Dynamics of Bright and Dark Excitons in 2H-WSe$_2$}}

\author{Mauro Fanciulli}
\email{mauro.fanciulli@u-cergy.fr}
\affiliation{Laboratoire de Physique des Matériaux et Surfaces, CY Cergy Paris Université, 95031 Cergy-Pontoise, France}
\affiliation{Université Paris-Saclay, CEA, CNRS, LIDYL, Gif-sur-Yvette, 91191, France}

\author{David Bresteau}
\affiliation{Université Paris-Saclay, CEA, CNRS, LIDYL, Gif-sur-Yvette, 91191, France}

\author{Jérome Gaudin}
\affiliation{Universit\'e de Bordeaux - CNRS - CEA, CELIA, UMR5107, F33405 Talence, France}

\author{Shuo Dong}
\affiliation{Beijing National Laboratory for Condensed Matter Physics, Institute of Physics, Chinese Academy of Sciences, Beijing 100190, China}

\author{Romain Géneaux}
\affiliation{Université Paris-Saclay, CEA, CNRS, LIDYL, Gif-sur-Yvette, 91191, France}

\author{Thierry Ruchon}
\affiliation{Université Paris-Saclay, CEA, CNRS, LIDYL, Gif-sur-Yvette, 91191, France}

\author{Olivier Tcherbakoff}
\affiliation{Université Paris-Saclay, CEA, CNRS, LIDYL, Gif-sur-Yvette, 91191, France}

\author{Ján Minár}
\affiliation{University of West Bohemia, New Technologies Research Centre, 301 00 Plzeň, Czech Republic}

\author{Olivier Heckmann}
\affiliation{Laboratoire de Physique des Matériaux et Surfaces, CY Cergy Paris Université, 95031 Cergy-Pontoise, France}
\affiliation{Université Paris-Saclay, CEA, CNRS, LIDYL, Gif-sur-Yvette, 91191, France}

\author{Maria Christine Richter}
\affiliation{Laboratoire de Physique des Matériaux et Surfaces, CY Cergy Paris Université, 95031 Cergy-Pontoise, France}
\affiliation{Université Paris-Saclay, CEA, CNRS, LIDYL, Gif-sur-Yvette, 91191, France}

\author{Karol Hricovini}
\email{karol.hricovini@u-cergy.fr}
\affiliation{Laboratoire de Physique des Matériaux et Surfaces, CY Cergy Paris Université, 95031 Cergy-Pontoise, France}
\affiliation{Université Paris-Saclay, CEA, CNRS, LIDYL, Gif-sur-Yvette, 91191, France}

\author{Samuel Beaulieu}
\email{samuel.beaulieu@u-bordeaux.fr}
\affiliation{Universit\'e de Bordeaux - CNRS - CEA, CELIA, UMR5107, F33405 Talence, France}

\begin{abstract}
We performed spin-, time- and angle-resolved extreme ultraviolet photoemission spectroscopy (STARPES) of excitons prepared by photoexcitation of inversion-symmetric 2H-WSe$_2$ with circularly polarized light. The very short probing depth of XUV photoemission permits selective measurement of photoelectrons originating from the top-most WSe$_2$ layer, allowing for direct measurement of hidden spin polarization of bright and momentum-forbidden dark excitons. Our results reveal efficient chiroptical control of bright excitons' hidden spin polarization. 
Following optical photoexcitation, intervalley scattering between nonequivalent K-K' valleys leads to a decay of bright excitons' hidden spin polarization. Conversely, the ultrafast formation of momentum-forbidden dark excitons acts as a local spin polarization reservoir, which could be used for spin injection in van der Waals heterostructures involving multilayer transition metal dichalcogenides.
\end{abstract}
\date{\today}
\maketitle

\begin{figure}[t]
\begin{center}
\includegraphics[width=8.5 cm,keepaspectratio=true]{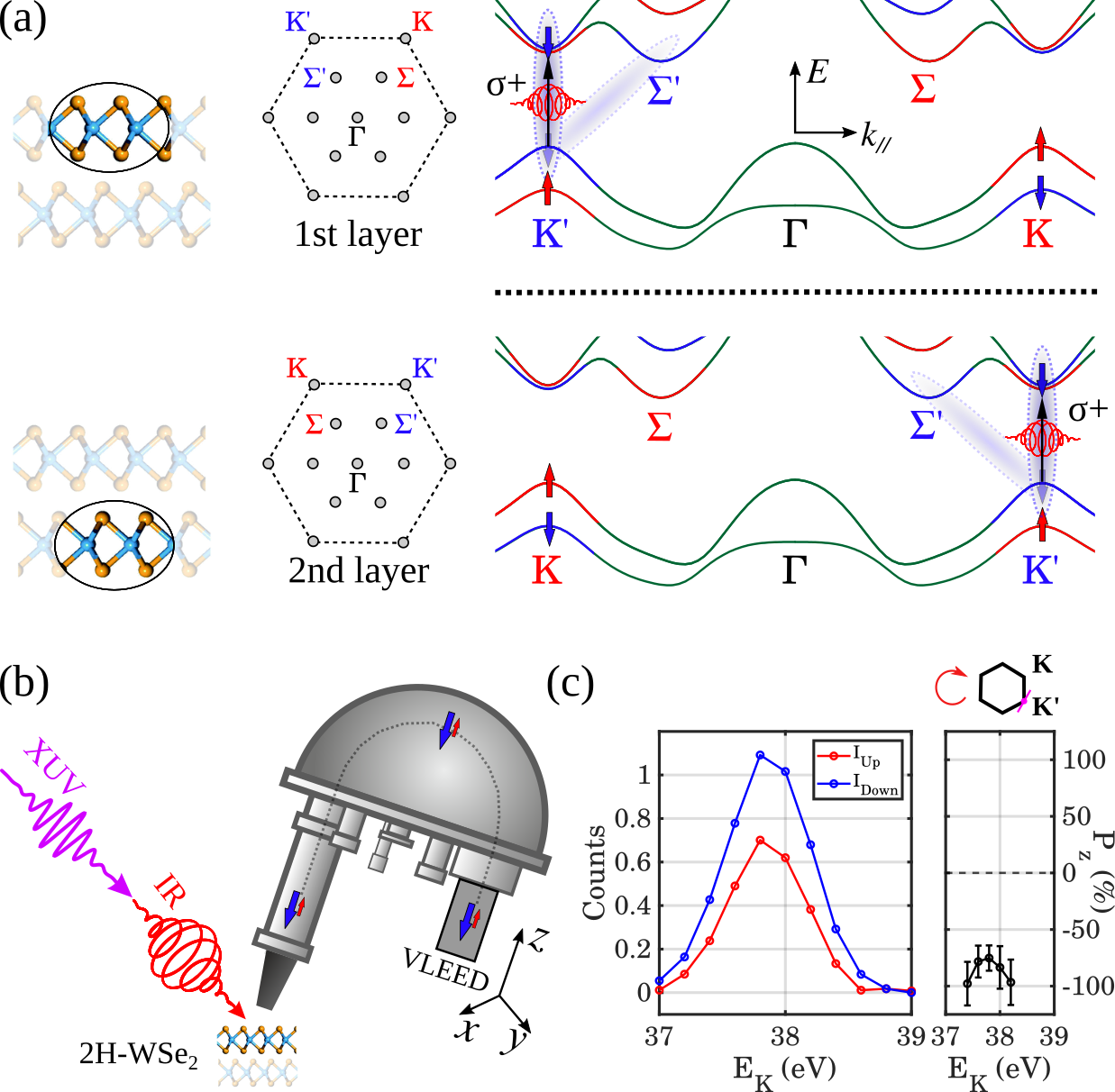}
\caption{\textbf{Hidden spin texture in 2H-WSe$_2$ and experimental methodology}: \textbf{(a)} Schematic of crystal structure, associated Brillouin zone and band structure for two adjacent layers in 2H-WSe$_2$. Valley-dependent chiroptical selection rules within each layer are visualized by a black arrow, bright and dark excitons are represented as shaded areas, and the spin texture is shown with red/blue color code. \textbf{(b)} Experimental setup: a circularly polarized infrared pulse (IR - 800~nm/1.55~eV) photoexcites 2H-WSe$_2$, while a linearly p-polarized extreme ultraviolet probe pulse (XUV - 34.78~nm/35.65~eV) with a delay $\Delta$t ejects photoelectrons mainly from the first layer, which are collected by the hemispherical analyzer with spin resolution. x and y axes are each at 45° from the out-of-the-page direction. \textbf{(c)} Spin-resolved ($z$ axis) EDCs of photoelectrons ejected from bright excitonic states in the majority K' valley ($\mathrm{I_{Up}}$ in red and $\mathrm{I_{Down}}$ in blue) and associated spin polarization (in black, right subpanel) at $\Delta$t=0~fs.}
\label{fig1}
\end{center}
\end{figure}

Spin-valley locking emerges in solids with broken inversion symmetry and strong spin-orbit coupling. This leads to peculiar momentum-dependent spin and orbital textures. Transition metal dichalcogenides (TMDC) are emblematic two-dimensional materials where this spin-valley locking leads to distinctive optical selection rules when using circularly polarized light, allowing for the generation of spin- and valley-polarized excitons~\cite{Mak12, Zeng12, Xiao12}. These concepts are at the foundation of spin-~\cite{Sierra21} and valleytronics~\cite{Schaibley16}. In bulk-TMDC of 2H polytype (e.g. 2H-WSe$_2$), adjacent layers are rotated by 180$^{\circ}$ with respect to each other, leading to opposite and alternating local momentum-space spin textures between neighboring layers (see Fig.~\ref{fig1}). This peculiar layered structure naturally introduces the concept of "hidden" spin texture~\cite{Zhang14}, which exists within each layer but vanishes in bulk, i.e. when the inversion-symmetry of the crystal is restored. TMDC hosts a great variety of so-called "hidden" properties, such as hidden orbital angular momentum and Berry curvature~\cite{Cho18}, intrinsic circularly polarized photoluminescence~\cite{Liu15}, spin-layer polarization~\cite{Guimaraes18}, and unconventional superconductivity~\cite{Liu17s}. Owing to the sub-monolayer inelastic mean free path of outgoing photoelectrons, the valence band's hidden spin texture of bulk-TMDC could be measured using extreme ultraviolet (XUV) spin- and angle-resolved photoemission spectroscopy (SARPES) \cite{Riley14, Razzoli17, Tu20}. As spintronic devices’ functionality arises in out-of-equilibrium states of matter, one very appealing route would be to extend this measurement methodology to the investigation of ultrafast hidden spin polarization dynamics of excited states in these layered van der Waals materials. 

The free carrier and exciton dynamics in 2H-WSe$_2$ have been extensively investigated using time- and angle-resolved photoemission spectroscopy (TR-ARPES)~\cite{Bertoni16, Liu17, Sie19, Maklar20, Dong21}, but, up to now, without spin resolution. It was shown that a near-resonant (800~nm/1.55~eV) pump pulse creates a coherent excitonic polarization, which dephases into an optically bright exciton population in less than 20 femtoseconds~\cite{Dong21}. These short-lived bright excitons subsequently relax through different intervalley scattering channels. A possible channel is scattering-backscattering between inequivalent K-K' valleys. The conduction band minimum spin-orbit-splitting at K and K' is only a few tens of meV~\cite{Roldan14}, thus K-K' scattering events are reversible and can either be mediated by intervalley electron-hole exchange~\cite{Maialle93, Schmidt16}, a process involving spin flip or be assisted by phonons~\cite{Selig16}, a spin-preserving process. K-K' intervalley scattering has been shown to be responsible for the rapid decay of hidden valley polarization~\cite{Bertoni16}, initially prepared using a circularly polarized pump pulse. Another possible relaxation channel is K-$\Sigma$ intervalley scattering, leading to the formation of momentum-forbidden dark excitons, with electron and hole residing at $\Sigma$ and K valleys, respectively~\cite{Lindlau18, Madeo20, Dong21}. These dark excitons are long-lived (tens of picoseconds)~\cite{Bertoni16}, due to their momentum-indirect nature. Surprisingly, following K-$\Sigma$ intervalley scattering, no hidden valley polarization was observed at $\Sigma$, despite the initial valley polarization in the K-K' valleys~\cite{Bertoni16}. It was argued that as states at $\Sigma$ are strongly delocalized in the direction across the layers, they are equally filled by scattering from K valleys in one layer and from K' in adjacent layers, washing out the valley- and layer-polarized nature of initial excited states at K-K' on an ultrafast timescale. These observations leave many open questions related to ultrafast dynamics of spin-polarized excitons in layered TMDCs, e.g.: Does hidden spin polarization of excitons survive intervalley scattering between adjacent K-K' valleys? Does bright excitons' initial hidden spin polarization remain upon the formation of dark excitons? Answering these questions is fundamental for designing spintronic device concepts based on multilayer TMDCs. However, directly accessing hidden spin polarization of TMDCs' excitons has not yet been demonstrated, mainly because of the famously challenging task of simultaneously combining time- and spin-resolution in ARPES, which are both extremely time-consuming. Indeed, while most successful attempts to combine time- and spin-ARPES were based on UV-Vis photoemission~\cite{Scholl97, Cinchetti06, Weber11, Cacho15, Jozwiak16, Sanchez16, Battiato18, Mori23}, this approach does not allow accessing large parallel momentum, which makes it blind to TMDCs' bright excitons, which are located at the Brillouin zone boundary. With the recent development of XUV STARPES using high-order harmonic sources~\cite{Plotzing16, Eich17, Nie19, Fanciulli20}, investigation of ultrafast exciton's spin-polarization dynamics in TMDCs can now be tackled. 

In this letter, we report the first spin-, time- and angle-resolved XUV photoemission spectroscopy of excitons in TMDC (here 2H-WSe$_2$). Our results demonstrate chiroptical control of bright excitons' hidden spin polarization, its decay upon intervalley scattering between adjacent K-K' valleys, and long-lived dark excitons with strong hidden spin polarization. 

The experiments were performed using the narrowband mode of the FAB10 beamline at the Attolab facility (CEA Saclay)~\cite{Bresteau23}. In a nutshell, we used a 10~kHz amplified Ti:Sa (1.55~eV) laser system delivering up to 2~mJ with a full width at half maximum (FWHM) duration of $\sim$23~fs. We split the beam into two arms: in the probe arm, a few hundred microjoules are focused in an argon gas jet to produce a broad spectrum of odd harmonics of the driving laser extending up to 50~eV, through high-order harmonic generation (HHG)~\cite{McPherson87, Ferray88}. A time-preserving monochromator is used to select a single harmonic (here the 23rd harmonic, 35.65~eV, $\sim$250~meV FWHM)~\cite{Poletto07} with linear (p-) polarization. The XUV pulse duration is estimated to be around 30~fs. In the pump arm, we used a polarization-tunable IR (1.55~eV) pulse, which is near the bright A-exciton resonance of 2H-WSe$_2$~\cite{Dong21}. The IR pump and XUV probe pulses are non-collinearly recombined onto the sample [Fig.\ref{fig1}(b)]. The pump fluence is estimated to be $\sim$ 1.9 $\mathrm{mJ/cm^2}$, which is very similar to the one used by Dong~\textit{et al.}~\cite{Dong21}, where clear photoemission signatures of bright excitons formation in 2H-WSe$_2$ were reported. The commercially available (HQ Graphene) bulk 2H-WSe$_2$ single crystal was cleaved at a base pressure of $\sim$2x10$^{-10}$~mbar. The measurements were performed at room temperature. The photoemission endstation comprises a hemispherical analyzer (SPECS PHOIBOS 150) and a 3D spin detector (Focus FERRUM)~\cite{Escher11}, based on very-low energy electron diffraction (VLEED). This detection scheme allows extracting the energy-resolved spin polarization along the three spin quantization axes in the detector reference frame, as shown in Fig.\ref{fig1}(b). An example of the measured spin polarization on the $z$ quantization axis ($\mathrm{P_z}$) of photoelectrons ejected from bright excitonic states at the majority K' valley (at the pump-probe overlap i.e. $\Delta$t=0~fs and for $\pm$7$^{\circ}$ ejection angles) is shown in Fig.~\ref{fig1}(c) (spin polarization data points for photoemission intensity smaller than 20$\%$ of the peak intensity are not shown). 

\begin{figure}[t]
\begin{center}
\includegraphics[width=8.5 cm,keepaspectratio=true]{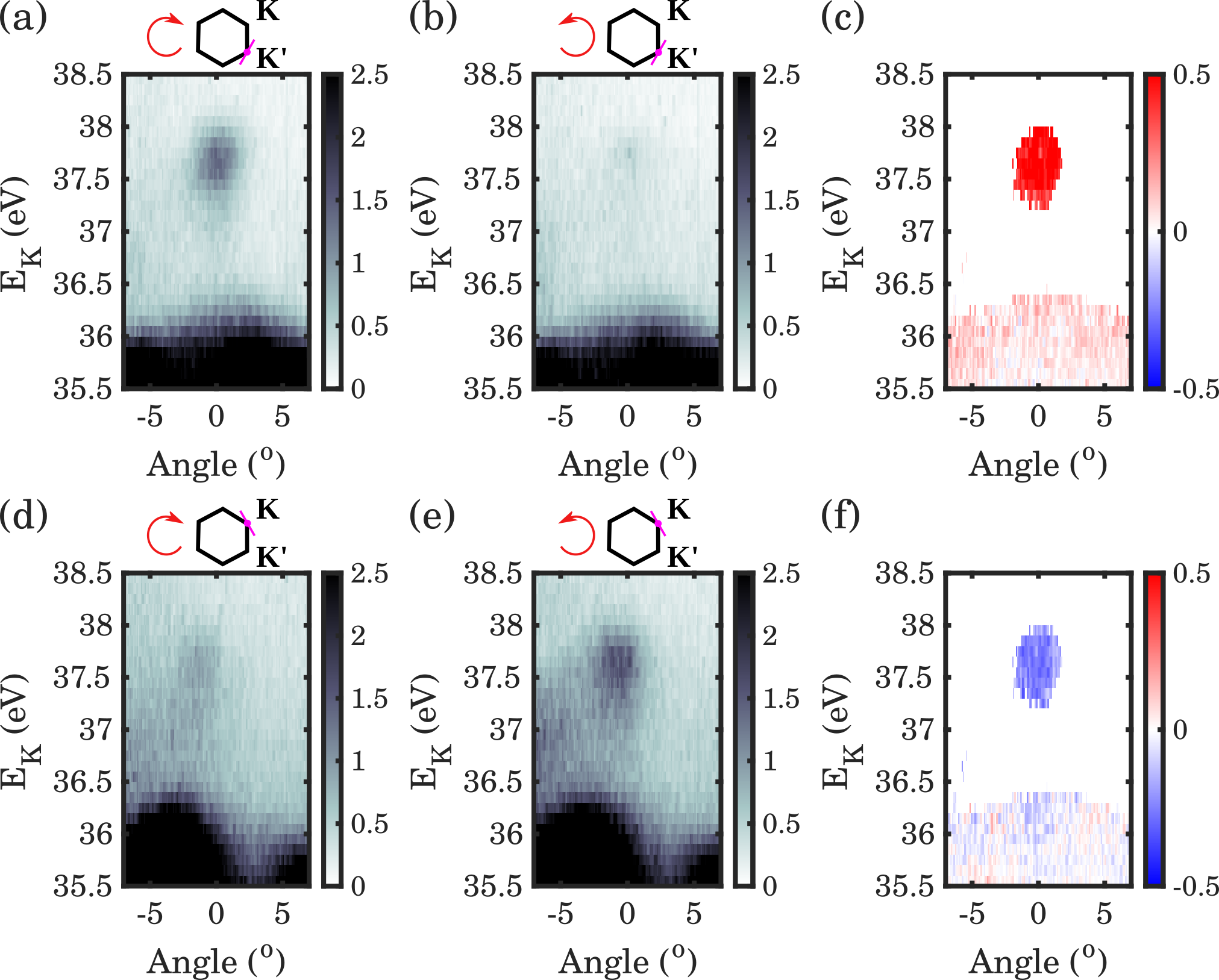}
\caption{\textbf{Spin-integrated circular dichroism emerging from valley-resolved chiroptical selection rules}: \textbf{(a)-(b)} and \textbf{(d)-(e)} Energy- and angle-resolved photoemission signal from bright excitons (K and K' valleys), for different pump pulse helicities as depicted on top of each panel. The magenta lines indicate the analyzer's slit direction. \textbf{(c),(f)} Valley-resolved circular dichroism associated with the photoemission intensities shown in \textbf{(a)-(b)} and  \textbf{(d)-(e)}, respectively. Circular dichroism data points for photoemission intensity smaller than 5$\%$ of the peak intensity are not shown.}
\label{fig2}
\end{center}
\end{figure}

We first investigate the spin-integrated bright exciton populations in K [Fig.~\ref{fig2} (a)-(b)] and K' Fig.~\ref{fig2} (d)-(e)] valleys after near-resonant photoexcitation with right and left circularly polarized light ($\sigma^+$ and $\sigma^-$), around the pump-probe overlap ($\Delta t$=0~fs). To experimentally swap the interrogated valley pseudospin index (K-K'), we azimuthally rotate the crystal by 60$^{\circ}$, which leaves all other experimental geometry parameters (e.g. angle of incidence) unchanged~\cite{Beaulieu20-2}. The photoemission intensity suppression in the valence band at around 2.5$^\circ$ from K [Fig.~\ref{fig2} (d)-(e)] is due to multiple orbitals interference effect, which has been discussed elsewhere~\cite{Rostami19, Beaulieu20-2}. The circular dichroism at K or K' ($\mathrm{CD}_{K/K'}$) is obtained by taking the normalized difference of the energy- and momentum-resolved signal at a given valley for different light helicity, i.e. $\mathrm{CD}_{K/K'} = [I^{\sigma^+}_{K/K'} - I^{\sigma^-}_{K/K'}]/[I^{\sigma^+}_{K/K'} + I^{\sigma^-}_{K/K'}]$ (Fig.~\ref{fig2}(c),(f)). We find a relatively strong CD exhibiting sign flip when changing the valley pseudospin index, indicating the initial preparation of bright excitons with strong hidden valley-polarization upon excitation with circularly polarized light. Our results agree with the experimental finding of Bertoni \textit{et al.}~\cite{Bertoni16} and are consistent with recent theoretical calculations that also reveal that these bright valley excitons formed upon absorption of circularly polarized light are chiral quasiparticles characterized by finite orbital angular momentum~\cite{Caruso22}. The different absolute values of CD at K and K' valleys might be due to a small pump-probe delay offset between the two measurements, also highlighted by the contribution of laser-assisted photoemission (LAPE)~\cite{Saathoff08} signal, stronger at K valley than at K' [i.e., the valence band (VB) replica at $\mathrm{E_{VB}+\hbar\omega_{IR}}$, well visible between 36.5-37.5~eV for negative emission angles in Fig.~\ref{fig2}(d)-(e)]. 

\begin{figure}[t]
\begin{center}
\includegraphics[width=8.5 cm,keepaspectratio=true]{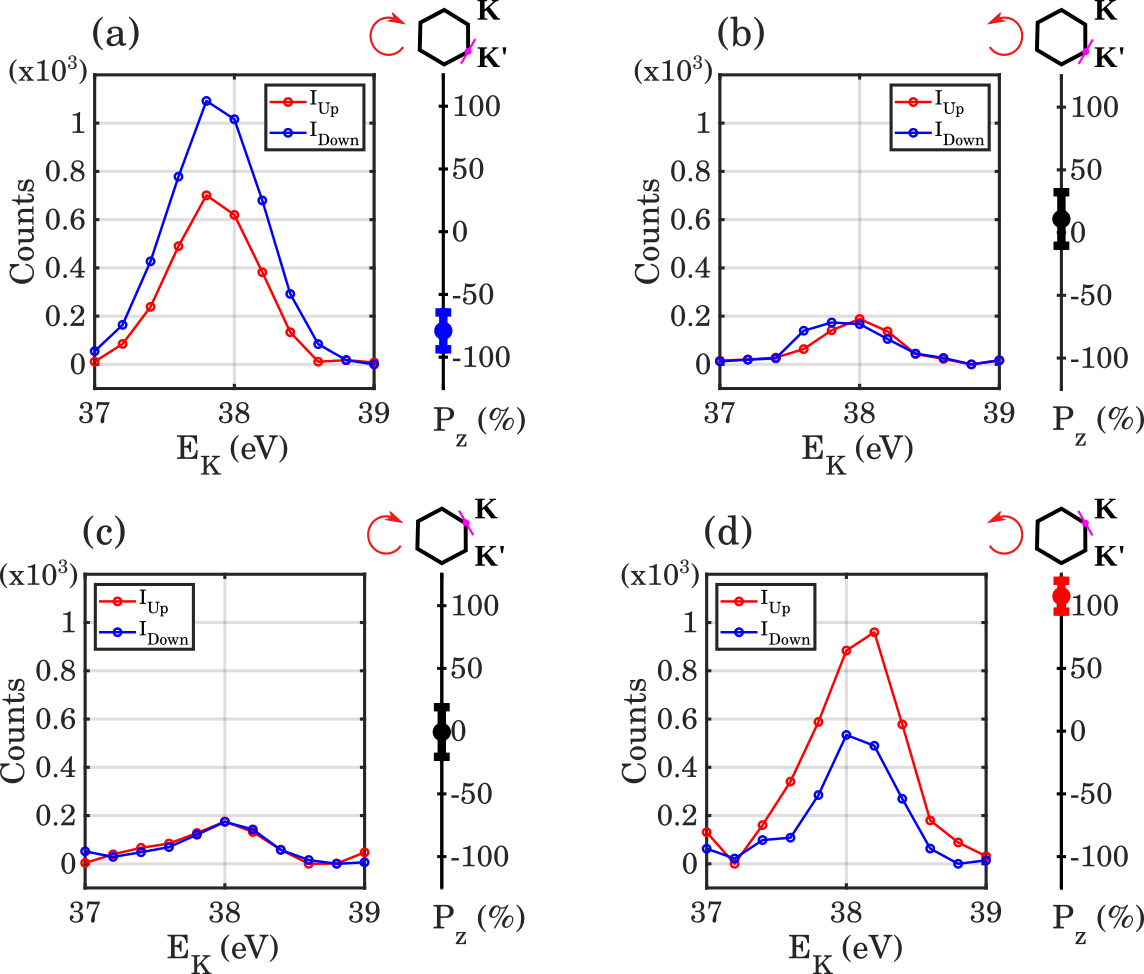}
\caption{\textbf{Valley- and helicity-resolved hidden spin polarization of bright excitons}: \textbf{(a)-(d)} Spin-resolved energy distribution curves ($\mathrm{I_{Up}}$ in red and $\mathrm{I_{Down}}$ in blue) of photoelectron ejected from bright excitonic states at both K' [\textbf{(a)-(b)}] and K [\textbf{(c)-(d)}] valleys and associated spin polarization (right subpanel), for both pump pulse helicities, at the pump-probe overlap. The pump helicities and valley indexes are depicted on top of each panel.}
\label{fig3}
\end{center}
\end{figure}

After looking at the hidden valley polarization induced by chiroptical selection rules, we investigate the hidden spin polarization of photoelectrons emerging from bright excitons in minority and majority valleys (Fig.~\ref{fig3}) at time zero, for both pump pulse helicities. Considering our photon energy, photoelectrons from the K/K' valleys are ejected towards the analyzer entrance slit with an angle of $25^\circ$ from the sample surface normal. Similarly to what has been measured for the valence band~\cite{Riley14}, we expect excitons' spin-polarization to be out-of-plane. Since we measure a vanishing spin polarization along the $x$ and $y$ quantization axes (see Supplemental Material \cite{SOM}), we consider only the $P_z$ spin-polarization component. This $P_z$ spin-polarization (detector frame) is strongly representative of the out-of-plane spin-polarization component (sample frame) due to the small angle between the surface normal and the detector axis. The spin polarization is obtained as $P_z=1/S\cdot\left(I_{Up}-I_{Down}\right)/\left(I_{Up}+I_{Down}\right)$, with $S$=0.29 the Sherman function~\cite{Escher11}, which takes into account the calibration of the spin detector. The reported values of $P_z$ are obtained by averaging the signal in a $\pm$200~meV energy interval around the energy distribution curve (EDC) peak and after exponential background subtraction for both spin channels $I_{Up, Down}$. The energy-resolved data are presented in \cite{SOM}. Spin-resolved measurements for each valley and polarization-state configurations were repeated 16 times and error bars represent the 95$\%$ confidence intervals calculated using Student’s statistics. The experimental data presented in Fig.~\ref{fig3} are obtained with a net acquisition time of 14 hours. As shown in Fig.~\ref{fig3}~(a) and (d), photoelectrons emerging from bright excitons in the majority valleys with both $\sigma^+$ [K' valley, Fig.~\ref{fig3}(a)] and $\sigma^-$ [K valley, Fig.~\ref{fig3}(d)] 
are almost fully spin-polarized (see Supplemental Material \cite{SOM} for a note on the absolute determination of light helicity, valley pseudospin index, and electron spin polarization). It is important to note that due to selection rules in photoemission, the measured spin polarization of outgoing photoelectrons cannot \textit{de facto} be linked to the initial state's spin polarization~\cite{dil19}. However, the sign reversal of the out-of-plane spin polarization with the valley pseudospin index allows us to safely link the measured photoelectron spin polarization and exciton's spin polarization, as it was concluded for the valence band~\cite{Riley14}. 

The spin polarization of photoelectrons emerging from excitons has the same sign as the ones emerging from the valence band top. Photoelectrons emerging from excitons in the minority valleys with both $\sigma^+$ [K valley, Fig.~\ref{fig3}(c)] and $\sigma^-$ [K' valley, Fig.~\ref{fig3}(b)] exhibit almost vanishing spin polarization. These observations indicate a balanced spin-up and spin-down population mixture of photoelectrons emerging from bright excitons population measured in the minority valley, which can originate from different microscopic scattering pathways leading to minority valley population, e.g. intervalley scattering driven by electron-hole exchange~\cite{Maialle93, Schmidt16}, or phonons \cite{Selig16}, or from imperfect light polarization-state due to the non-normal incidence angle on optics and sample.

\begin{figure}[t]
\begin{center}
\includegraphics[width=8.5 cm,keepaspectratio=true]{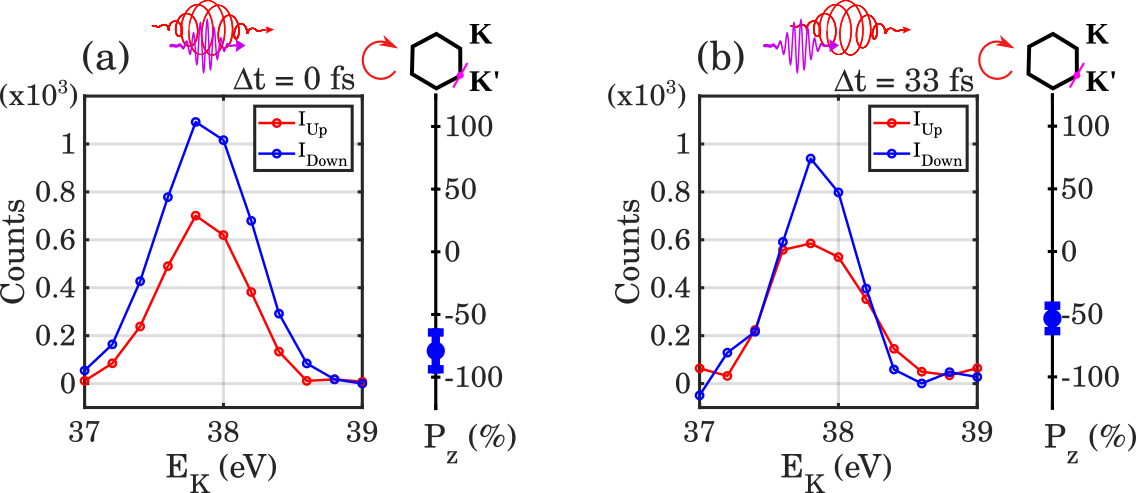}
\caption{\textbf{Ultrafast decay of the hidden spin polarization of bright excitonic states}: Spin-resolved EDCs of photoelectron ejected from bright excitonic states at K' valley ($\mathrm{I_{Up}}$ in red and $\mathrm{I_{Down}}$ in blue) and associated spin polarization (in black, right subpanel), \textbf{(a)} at pump-probe overlap, i.e. $\Delta$t=0~fs and \textbf{(b)} at $\Delta$t=33~fs.}
\label{fig4}
\end{center}
\end{figure}

\begin{figure}[t]
\begin{center}
\includegraphics[width=8.5 cm,keepaspectratio=true]{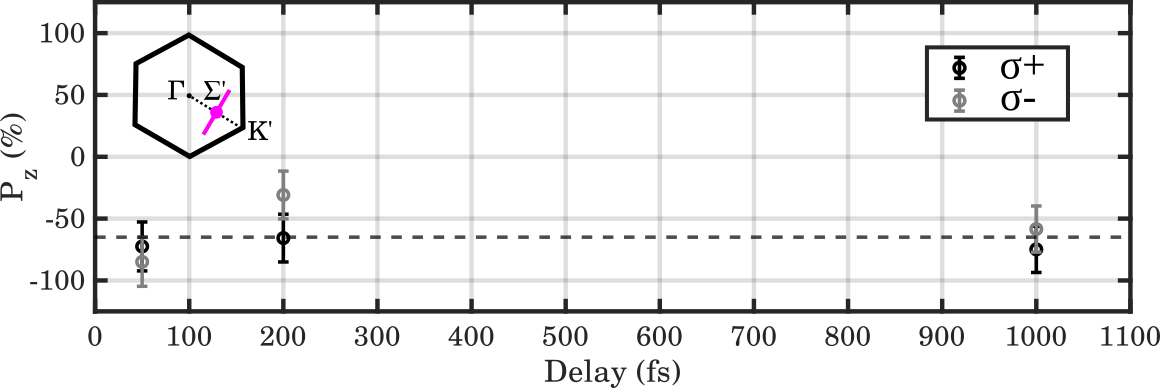}
\caption{\textbf{Hidden spin polarization dynamics of momentum-forbidden dark excitons}. The black and grey dots represent measured spin polarizations of photoelectrons ejected from dark excitonic states at $\Sigma$' valley, for three pump-probe delays (50~fs, 200~fs and 1000~fs) and for both pump pulse helicities ($\sigma^+$ in black, and $\sigma^-$, in grey). The dotted line represents the mean spin polarization value for both polarization and all delays.}
\label{fig5}
\end{center}
\end{figure}

We now turn our attention to the ultrafast (femtosecond) dynamics of the excitons' spin polarization initially prepared by a chiroptical transition. Valley polarization created upon circularly polarized excitation is known to decay on a sub-100~fs timescale, due to K-K' intervalley scattering [$\tau_{KK'}$=(60$\pm$30)~fs]~\cite{Bertoni16}. When changing the pump-probe delay from $\Delta t$=0~fs to $\Delta t$=33~fs, bright excitons' hidden spin polarization is found to decay from -79$\pm$15$\%$ to -53$\pm$10$\%$ [Fig.~\ref{fig4}]. This spin- and time-resolved measurement allows us to get additional insight into K-K' intervalley scattering.  Indeed, spin polarization decay can only
happen if both intervalley electron-hole exchange~\cite{Maialle93, Schmidt16}
and intervalley phonon-assisted population transfer contribute to scattering-backscattering between nonequivalent K-K’ valleys. Indeed, in a scenario where only intervalley electron-hole exchange would be active, each
scattering event between K-K’ (or vice-versa) would involve a spin-flip process, which could quench the valley polarization but would leaves spin polarization in each valley time-independent. The same situation occurs if only spin-preserving phonon-assisted scattering between K-K’ would be allowed. While it is not possible to extract some relative contribution from these two channels due to the time-consuming nature of these measurements (only two pump-probe delays), we can safely conclude that this bright excitons’ hidden spin polarization decay is due to a combined and reversible intervalley electron-hole exchange and phonon-assisted scattering between nonequivalent K-K’ valleys. One open question is related to bright excitons' spin polarization dynamics over longer timescales: does it completely vanishes, or does it saturates? Future STARPES investigations using high-repetition-rate beamlines would allow measuring additional pump-probe delays and resolve the complete temporal evolution of bright excitons spin polarization. 

In bulk 2H-WSe$_2$, global conduction band minima are located at $\Sigma$, which is located roughly halfway between $\Gamma$ and K. The single-particle band structure predicts a spin-orbit-splitting of almost 1~eV at the $\Sigma$ conduction band~\cite{Roldan14}. One of the dominant relaxation pathways for bright excitons is the formation of long-lived momentum-forbidden dark excitons, where electrons and holes reside at $\Sigma$ and K valley respectively~\cite{Dong21}. K-$\Sigma$ intervalley scattering has been reported to lead to a loss of valley- and layer-polarization. This observation was rationalized by the three-dimensional character of the states at $\Sigma$. It is still an open question whether or not the loss of valley and layer polarization is accompanied by a loss of spin polarization. It is thus of capital importance to track the dynamical evolution of the hidden spin polarization upon the formation of such dark excitons, to reveal if it is possible to harvest optically-induced initial spin-polarized bright excitons into long-lived dark excitonic states, for spintronics applications.

In Fig.~\ref{fig5}, we measured the hidden spin polarization of momentum-forbidden dark excitons at $\Sigma$' for three pump-probe delays (50~fs, 200~fs, and 1000~fs) and both pump pulse helicities. The spin polarization is negative for both pump helicities and all investigated pump-probe delays. The fact that spin polarization has the same sign for both helicity gives us strong insights into the scattering mechanisms involved in the creation of dark excitons. Indeed, this implies that the strong spin-orbit splitting at $\Sigma$ imposes a given final spin state for each scattering event leading to its population. Thus, despite the vanishing $\Sigma$ valley-polarization following K-$\Sigma$ intervalley scattering~\cite{Bertoni16}, momentum-forbidden dark excitons are locally (in reciprocal space, i.e. within each valley) spin-polarized. A picosecond after photoexcitation, hidden spin polarization has almost the same amplitude as at early time delay (50~fs), despite slightly smaller measured values at intermediate pump-probe delay (200~fs). Resolving the complete temporal evolution (tens of pump-probe delay, for both pump helicities) of (hidden) spin polarization dynamics would be highly desirable for elucidating more subtle spin relaxation mechanisms, but is not reachable using the current setup.  

Our results report efficient chiroptical control of excitons' hidden spin polarization in bulk 2H-WSe$_2$ and its ultrafast dynamics upon intervalley scattering. Our measurements reveal quasi-fully spin-polarized excitons in the majority valley upon photoexcitation with circularly polarized light. We find that subsequent K-K' intervalley scattering is due to two microscopic scattering channels, intervalley electron-hole exchange (spin-flip process) and intervalley phonon-assisted population transfer (spin-preserving process), leading to ultrafast spin polarization decay for bright excitons at K and K' valleys. Instead, the formation of momentum-forbidden dark excitons through K-$\Sigma$ intervalley scattering acts as a local momentum-space spin-reservoir. Indeed, despite the ultrafast decay of valley- and layer-polarization~\cite{Bertoni16} following photoexcitation, the strong spin-orbit splitting at $\Sigma$ allows the survival of the helicity-independent local hidden spin polarization for dark excitons. This long spin-polarization lifetime is desirable for spintronic applications. Our approach can be directly extended to a wide range of out-of-equilibrium spin dynamics of many-body quasiparticles in solids. In addition, combining this STARPES methodology with a polarization-tunable (circularly polarized) XUV probe pulse would allow accessing the orbital angular momentum and chirality of these optically-prepared spin-polarized excited states~\cite{Caruso22, Cho18, Schuler20}.  

\begin{acknowledgments}
\textbf{Acknowledgments} The laser system and the experimental setup are supported by the French “Investments for the Future” of the Agence Nationale pour la Recherche, Contracts No. 11-EQPX0005-ATTOLAB, and No.11-EQPX0034-PATRIMEX. The laser system is also supported by the Scientific Cooperation Foundation of Paris-Saclay University through the funding of the OPT2X research project (Lidex 2014), by the Île de France region through the Pulse-X project, and by the European Union’s Horizon. 
\end{acknowledgments}

%\bibliography{STARPES_WSe2} % Produces the bibliography via BibTeX.

\providecommand{\noopsort}[1]{}\providecommand{\singleletter}[1]{#1}%

\end{document}